\documentclass[aps,pra,10pt,a4paper,twocolumn]{revtex4-2}
\usepackage{graphicx} 
\usepackage{amsmath}

\newcommand{\lvec}{\boldsymbol{l}}
\newcommand{\rvec}{\boldsymbol{r}}
\newcommand{\svec}{\boldsymbol{s}}
\newcommand{\Rho}{\mathrm{P}}

\begin{document}

\title{Designing cultured tissue moulds using evolutionary strategies}
\author{Allison E. Andrews}
\affiliation{The Open University, Walton Hall, Milton Keynes, MK6 3AY, UK}
\author{Hugh Dickinson}
\affiliation{The Open University, Walton Hall, Milton Keynes, MK6 3AY, UK}
\author{James P. Hague}
\affiliation{The Open University, Walton Hall, Milton Keynes, MK6 3AY, UK}

\date{\today}

\begin{abstract}
    There is an unmet need for artificial intelligence techniques that can speed up the design of growth strategies for cultured tissues. Cultured tissue is increasingly important for a range of applications such as cultivated meat, pharmaceutical assays and regenerative medicine. In this paper, we introduce a method based around evolutionary strategies, machine learning and biophysical simulations that can be used to speed up the process of identifying new tissue growth strategies for these diverse applications. We demonstrate the method by designing tethering strategies to grow tissues containing various cell types with desirable properties such as high cellular alignment and uniform density.
\end{abstract}

\maketitle

\section{Introduction}

Simulation, machine learning and computational intelligence techniques are needed to assist in the design of growth strategies for cultured tissue with bespoke properties. Cultured tissue is highly important for a range applications such as cultivated meat \cite{benarye2019a}, pharmaceutical assays \cite{weinhart2019a,jensen2018a}, regenerative medicine \cite{bajaj2014a} and in general for the study of fundamental biology. In particular, the recent FDA Modernization Act 2.0 means that the medicines pipeline no longer requires animal testing, making 3D tissue cultures increasingly relevant to pharmaceutical development \cite{han2022}. Tethered moulds, 3D printing, scaffolds and spheroids can be used to grow cultured tissues \cite{bajaj2014a}. This paper considers tethered cell-laden hydrogels, which have been used for the growth of artificial tissues describing muscle \cite{capel2019a}, neural tissue \cite{georgiou2013}, tendons \cite{garvin2003a}, cornea \cite{mukhey2018} and for studying fibroblasts \cite{kostyuk2004}, and investigated for high-throughput screening \cite{capel2019a,wragg2019a} and pharmacological testing \cite{garvin2003a}.

A goal of tissue engineering is to mimic the self-organisation of cells in tissues to improve functionality. Specific organisations of cells and matrix are essential to the functionality many tissues, such as contraction in skeletal muscle \cite{grosberg2011a,shimizu2009}, the optical properties of the cornea \cite{gouveia2017a,mukhey2018}, the action of fibroblasts during healing \cite{werner2007a} and support of neurons by glial cells \cite{deumens2004a,georgiou2013}.

Cultured tissues self-organise in a bottom-up process, whereas the design of growth strategies (such as the tethered moulds) is top-down, meaning that traditional computer aided design is of limited use in the design process. Simulations can help to bridge the gap between top-down and bottom-up approaches. To develop rapid predictions, we have developed the RAPTOR approach to predicting the results of bottom-up self-organisation using a machine-learning approach \cite{andrews2023,andrews2025a}. Machine learning makes fast predictions, and individual predictions can be made in less than a second. This opens the possibility to automate the design process for the creation of 3D mould and scaffold design for specific tissue properties.

In this proof-of-concept study we demonstrate how evolutionary strategies can assist with the design of tethered moulds for cultured tissues with favourable properties that comply with manufacturing constraints. Maximizing regions of alignment while reducing misaligned areas can be important for tissue functionality. High tension in tethered tissues may lead to a risk of breakage, so designs should aim to reduce it. Furthermore, manufacturing constraints of the moulds may need to be added to the evolutionary strategies (e.g. to ensure that tethers are sufficiently wide and situated away from boundaries). Previously, we carried out a straightforward stochastic search based on randomly generated (but plausible) moulds to find candidate moulds with very high cell alignment and low internal tension \cite{hague2023a}. Standard designs of tethered moulds for e.g. tendons, muscle, glial tissue, fibroblast tissue and corneal tissue Refs.  \cite{capel2019a,wragg2019a,garvin2003a,orourke2017,orourke2015,mukhey2018,eastwood1998,east2010} can suffer from regions of misaligned tissue, especially around the tethers \cite{eastwood1998,georgiou2013}. 

The machine learning method used to predict tissue organisation (such that the fitness of candidate solutions can be determined) is based on the contractile network dipole orientation (CONDOR) model, which describes the self-organisation of cells interacting through the extra-cellular matrix. The configuration of the matrix is modelled as a contractile network of bonds on a face-center cubic lattice,
\begin{equation} 
E = \sum_{i<j} \frac{\kappa_{0}\bar{\kappa}_{ij}}{2}\left( |\lvec_{ij}| - l'_{ij}\right)^2,
\label{eqn:hamiltonian}
\end{equation}
with the action of cells on the matrix modelled by a change in the equilibrium length of bonds related to a change in direction of force dipoles representing cells,
\begin{equation}
l'_{ij} = l_{0}\left(1-\frac{\Delta}{2}\left(2-|\hat{\lvec}_{ij}\cdot\svec_{i}|^2 -|\hat{\lvec}_{ij}\cdot\svec_{j}|^2\right) \right).
\end{equation}
where $l_0$ is the equilibrium bond length and the cell-matrix interaction is $\Delta$, which is the parameter that controls the contractility of the tissue (i.e. a higher $\Delta$ represents a tissue where the cells cause a high level of contraction). The contractile network consists of springs with spring constant $\kappa_{ij} = \kappa_{0}\bar{\kappa}_{ij}$ where $\kappa_{0}$ is the nearest neighbour spring constant and $\bar{\kappa}_{ij}$ is a dimensionless spring constant. The displacement between cells $i$ and $j$ is denoted $\lvec_{ij}$. The dipole orientations of cells are denoted $\svec$ (i.e. the direction of $\svec$ represents the orientation of each cell) and the orientation vector of bonds as $\hat{\lvec}=\lvec/|\lvec|$. Specific values for spring constants are $\bar{\kappa}_{2n}=1$, $\bar{\kappa}_{3n}=0.5$, $\bar{\kappa}_{4n}=0.25$ ($\bar{\kappa}_{4n}/\bar{\kappa}_{3n}=0.5$), where $2n$ represents nearest neighbour, $3n$ next-nearest neighbour, and so on.

The configuration of cells and matrix associated with the minimum value of $E$ has been shown to provide a good match to the organisations of artificial tissues (such as glial, fibroblast and corneal tissues) where alignment is promoted \cite{hague2019a, hague2019a,andrews2023,hague2023a}. Different values of $\Delta$ and $\kappa$ are associated with different cells and ECMs. $\Delta$ parameterises the contraction of the matrix perpendicular to the cell orientation, and represents the remodelling and forces applied to the ECM by the cells. $\kappa$ values in the contractile network can be related to ECM properties such the various elastic module \cite{boal}. ECM in tissues provides an important framework for structure, maintenance, repair, tissue development and promotion of inter-cell communication \cite{kular2014a}, so while it is missing from many biomechanical models, its presence in the model is essential for a range of tissue properties.

The goal of this paper is to show proof-of-concept for evolutionary strategies (genetic algorithms) for tethered mould design. We use machine learning predictions of cell-matrix interactions in combination with evolutionary strategies. Tethered moulds with specific tissue patterns are designed, e.g. large regions of high cell alignment and uniform density. The machine learning approach is introduced in \cite{andrews2023,andrews2025a}. This Genetic algorithm (and) neural net (for) engineered tissue (GANNET) is used to design tethered moulds for optimally aligned cell arrangements and uniform cell densities in tissues. The use of evolutionary strategies to design tethered moulds using biophysical principles goes beyond previous work in this area. Evolutionary strategies allow for a fast and automatic design process before embarking upon time consuming and expensive growth and analysis of tissues to understand scaffolding strategies in the lab.

This paper is organised as follows: In Section \ref{sec:ga}, we introduce a genetic algorithm approach that works with RAPTOR to design tethered moulds. In Section \ref{sec:results} we design tethered moulds suitable for different tissue types and regions of alignment. We summarize, discuss the outlook and make conclusions in Section \ref{sec:conclusions}.

\section{Genetic algorithm}
\label{sec:ga}

\subsection{Algorithm overview}

Genetic algorithms (GA) are used here to find optimal mould designs. They are a form of optimisation algorithm that can be used to find solutions of complex or abstract problems. Used as a design tool, they constitute a form of artificial or computational intelligence. This section provides an overview of the algorithm used here, with specifics to be found in Secs. \ref{sec:encoding} to \ref{sec:fitness}.

Genetic algorithms are inspired by natural selection \cite{goldberg1989}. A very large sample of randomised candidate solutions is generated from encoding parameters and their fitness for purpose numerically evaluated and ranked. Analogues of genetic crossover and mutation are then applied to create new members of the population, with less optimal candidates removed. Repetition of this process leads to candidate solutions with increasingly optimal properties.

The genetic algorithm used here is initialised with 1000 candidate solutions with random parameter values as a starting population (See Sec. \ref{sec:initial}). Each solution is used to generate a mould design, which is subsequently converted into a two channel image (see Sec. \ref{sec:encoding}). Designs are then used as inputs for the RAPTOR model to rapidly predict tissue properties before the fitness functions of all initial mould designs are evaluated (see Sec. \ref{sec:fitness}).

New candidates for successive generations are produced via crossover operations, in which parameters from existing pairs of candidates are combined (see Sec. \ref{sec:crossover}), in addition to mutation of existing parameters (see Sec. \ref{sec:mutation}); hence the genetic part of the algorithm. Candidate pairs used in crossovers are selected via a tournament selection method with a tournament size of $2$, i.e. random pairs of candidates are compared against each other in terms of fitness value until two high-scoring (though not necessarily the most highest scoring) remain. Candidates have a fixed probability to undergo mutation (the mutation rate) that we set at $0.2$. The resulting candidates are converted into mould designs and their fitness functions are evaluated.

The new candidates produced by the crossover and mutation are added to the population. Following this, the lowest two ranking solutions are then removed, returning the population size to 1000. This process is repeated for new pairs of candidates until a total of $15000$ evaluations have been completed, equivalent to cycling through the population 15 times. 

The process of generating a single mould and making a machine learning prediction using RAPTOR takes less than a second, which is what enables the use of optimisation routines such as the genetic algorithm. We make use of the Inspyred library \cite{tonda2020}. For both the specific cases of maximum alignment and uniform density that we test here, the entire process in takes approximately 45 minutes to complete in each case.

\subsection{Mould encoding}\label{sec:encoding}

Candidate solutions are encoded as a set of parameter values which can be used to generate a mould shape and tether layout, using a methodology similar to that described in \citet{andrews2023}. These parameter values include (1) positions of the mould vertices (2) positions and radial sizes of the (circular) tethers and (3) parameters to control the rounding of convex and concave vertices in the final mould shape; candidate solutions use up to a total of $37$ continuous parameter values. We also define an additional set of binary parameters that act as switches to control whether individual mould vertices or tethers are used in the mould generation, allowing the total number appearing in final moulds to vary for both.

We generate mould designs on a square canvas of side $2a$. In the following, unitless dimensions, $x/a$ and $y/a \in (-1,1)$ are used for coordinates and areas are in units of $A$. Mould shapes and tether configurations are generated with vertices and tethers mirrored along $x=0$ and $y=0$.

Numerical bounds are applied to all parameter values to prevent unfeasible or impractical moulds from being generated. The bounding values for all parameters are shown in Table~\ref{table:parameter_limits}. We impose additional limits on vertex and tether coordinates such that their radial distance from the point $(0,0)$ does not exceed the upper bound as specified in Table~\ref{table:parameter_limits}.

\begin{table}
\centering
\caption{Mould generation parameter limits. Limits for coordinates apply to both the x and y axes}
\begin{tabular}{ c|c|c } 
Parameter & Lower bound & Upper bound \\
\hline
Vertex coordinates & $0.1$ & $0.9$ \\
Tether coordinates & $0.0$ & $0.85$ \\
Tether size & $0.02$ & $0.1$ \\
Convex rounding & $0.0$ & $0.1$ \\
Concave rounding & $0.0$ & $0.1$ \\
\hline
\end{tabular}

\label{table:parameter_limits}
\end{table}

Mould designs are generated from candidate solutions using functions within the python package shapely, following a routine similar to that described in \cite{andrews2023}. This routine firstly involves generating a shape boundary using the mould vertices switched `on', with the edges of the resulting shape subsequently rounded using the convex and concave rounding values defined in the candidate solution. Following this, tethers are generated within the mould. Tethers that lie either fully or partially outside the bounds of the mould shape are not generated, regardless of whether switched `on'. Once a candidate has been used to generate a mould design, the design is converted into an image format that can be used as an input for our RAPTOR prediction model.

\subsection{Initial population}
\label{sec:initial}

The initial population is generated from a large and varied sample of possible parameter values.

Initial mould vertex and tether coordinate pairs are randomly generated in a polar coordinate system. The radial $r$ values for these are generated uniformly within their respective bounds as listed in Table~\ref{table:parameter_limits}. To ensure the mould boundary path created by vertices does not cross itself for any candidate in the initial population (which would cause mould generation to fail), each vertex is generated with a fixed and separate range of $\theta$ values. For $N_{v}$ vertices, vertex $n$ sits in the angular range $(n-1)\pi/2N_{v} \leq \theta \leq n\pi/2N_{v}$, where $\theta$ is in radians, overall comprising a $0$ to $\pi/2$ radian range. This limitation is not placed on tether coordinates, which are instead randomly generated with $0 \leq \theta \leq \pi/2$. We use a total number of vertices $N_{v}=10$, and total number of tethers $N_{t}=5$.

Mould vertex and tether switches are set to true or false with equal probability, with the exception of the first and last mould vertices and first tether which are always set to true (and hence, always present). The remaining continuous parameter values of candidates, including tether size, convex rounding and concave rounding, are generated uniformly using the ranges shown in Table~\ref{table:parameter_limits}.

\subsection{Crossover}
\label{sec:crossover}

Crossover is an essential part of the genetic algorithm that allows the encoding from two parents to be combined to create two new offspring encodings. Crossover operations are defined for tethers, vertices and rounding parameters. 

\begin{figure}
\centering
{
\includegraphics[width=0.47\textwidth]{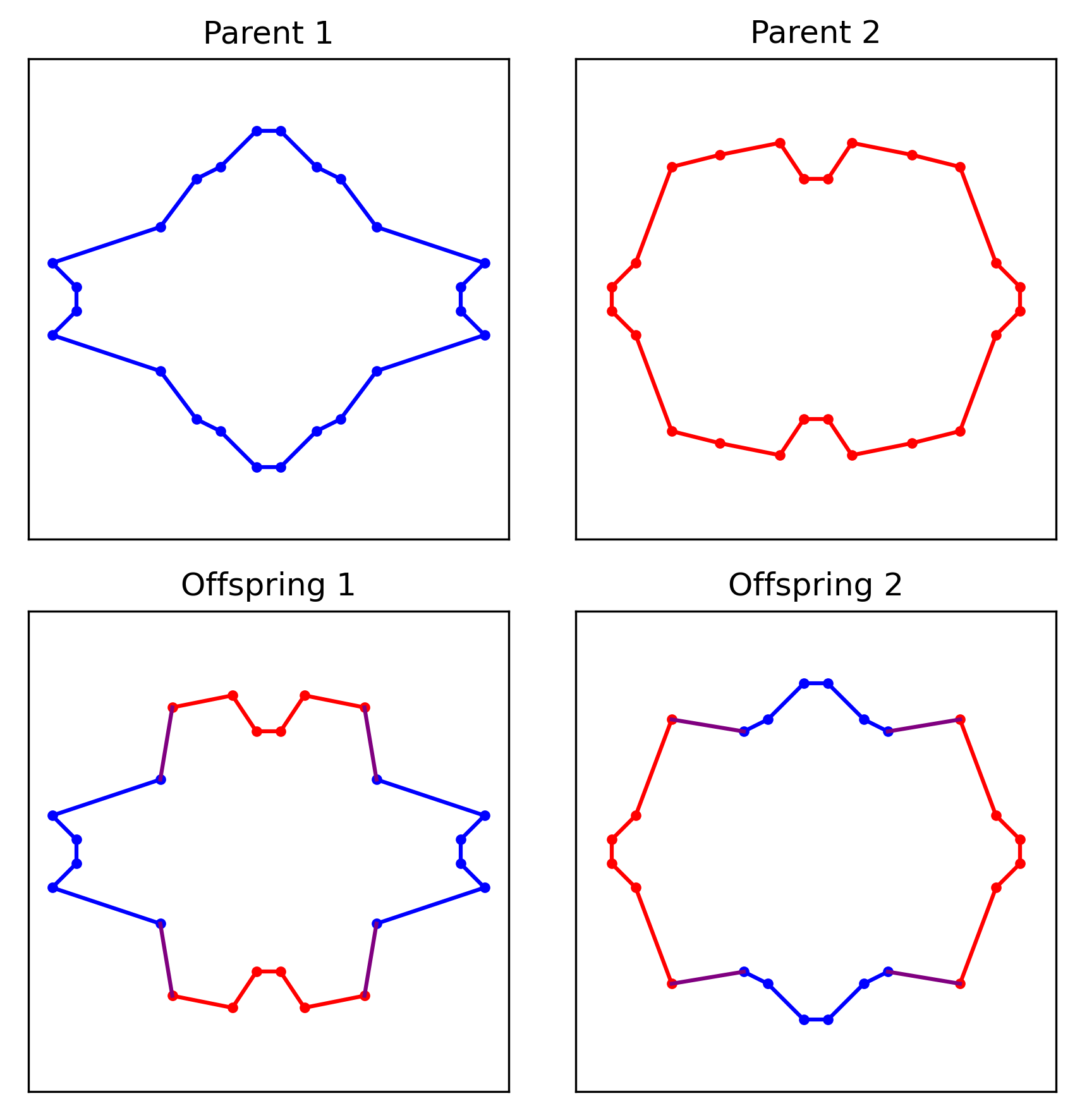}
\caption{An example of how a crossover of mould vertices produces two new mould shapes with features of both parent candidate solutions.}
\label{fig:crossover_example_1}
}
\end{figure}

To preserve mould geometry, the crossover routine swaps sequences of vertices of two parent candidates at a crossover point (including those switched off). The random position includes both the first and last coordinates, hence all coordinates can be completely swapped or not swapped at all in a crossover. An example of a mould vertex crossover is shown in Fig.~\ref{fig:crossover_example_1}.

We apply the same procedure to tether coordinates. A crossover position for the sequence of tether tuples is randomly determined. The final two parameters in candidate solutions, controlling the convex rounding and concave rounding of the entire mould shape, each have a separate 50\% chance of being swapped in offspring encodings.

When a crossover operation is carried out, all of the routines for crossing mould vertex, tether and shape parameter data are applied. The genetic algorithm allows for the total crossover rate, i.e. the probability of these algorithms being applied and new offspring solutions created, to be adjusted; for all cases in this work we have set this crossover rate to be 100\%, hence new offspring candidates are generated from all existing candidates in every successive generation.

\subsection{Mutation}
\label{sec:mutation}

Mutation modifies candidate solutions by applying random adjustments to parameters of existing solutions. The individual parameters, including each mould vertex tuple, tether tuple, as well as the pair of shape rounding parameters (which are treated as a single component), are given a set probability of mutation. The mutation rate is set at 20\% so that typically several (but not all) components of the candidate solution are changed.

For the continuous parameter values in candidates (i.e. coordinates of vertices, widths and positions of tethers, and rounding parameters) two types of mutation are applied. (1) A local mutation, which involves applying a small offset to a parameter value. The value offset is randomly determined via a Gaussian random variate with standard deviation of 10\% of the parameter limits shown in Table~\ref{table:parameter_limits}. (2) a global mutation where the parameter value is randomly set to any possible value within the radial limits discussed in Sec. \ref{sec:initial}. One of these two types of mutation is selected with equal probability.

For local mutation operations, it is possible in a small number of cases for a mould or tether coordinate to be pushed slightly outside of the designated coordinate bounds. When this occurs, the coordinate's radial position is readjusted to place it back within the coordinate bounds; we do not make use of periodic boundaries as they would result in a large parameter adjustment. Similarly, random local offsets applied to tether sizes or rounding are also readjusted to prevent the values falling outside their bounds. We note that this makes these random local offsets non-gaussian for parameter values near the extremes, but this did not cause any obvious biases to extremal values in the designs (as seen in Fig.~\ref{fig:max_alignment_low_delta_mould_examples}, \ref{fig:max_alignment_high_delta_mould_examples} and \ref{fig:uniform_density_mould_examples}).

For the boolean mould and tether switches, the probability of state change during mutation is 15\%. Switches for first and last mould vertex and first tether are unchanged, as to always ensure that a mould shape is generated and with at least one mirrored set of tethers.

\subsection{Fitness functions}
\label{sec:fitness}

We calculate fitness functions using RAPTOR predictions \cite{andrews2025a}; as already mentioned candidate solutions are converted into appropriate inputs for this machine learning model. It is possible for mould generation to fail in rare circumstances, for example cases where the arrangement of mould vertices creates a disjointed mould. Where this occurs, the fitness is automatically downrated to give the candidate the lowest ranking of the population.

We study two distinct optimisation objectives for generated moulds: (1) moulds that will produce maximised alignment in cells; and (2) moulds that produce an area of uniform density. We also make use of multiple penalties to remove unfeasible or impractical moulds. Calculations for fitness functions as well as penalties used for each differ.

\subsubsection{Maximum Alignment}

Our first optimisation goal is to produce a mould which maximises the alignment of cells. For this, we make use of the RAPTOR predictions for the $x$ component of the cell orientation field, $S_{x}^{2}$, and cell density $\Rho$. The fitness value is calculated from the mean orientation, $\langle \svec_{x} \rangle$ which is the position average of $S_{x}(\rvec)^{2}/\Rho(\rvec)$ \citep{andrews2023}. The resulting value lies in a range between $0$ and $1$, with $1$ indicating complete alignment. This value is independent of the mould area, allowing for comparison between different mould designs. Three penalties are applied to ensure that the designs incorporate specific useful features.

With the measure of alignment being independent of mould area, the first penalty is used to prevent small moulds from being generated. We assign a minimum mould area $A_{\rm min}$, defined as a fraction of the total canvas area, $A$ ($A=256^{2}$ pixels), with a typical value $A_{\rm min}/A=0.25$. This is set up as a soft limit; moulds can be generated with an area below this limit, but incur an increasing penalty to the fitness value. This area penalty,
\begin{equation}
p_{A} = C_{A}\,{\rm max}\left(1-\frac{A_{\mathrm{mould}}}{A_{\mathrm{min}}},0\right)
\label{eqn:areapenalty}
\end{equation}
where $A_{\mathrm{mould}}$ is the area of the mould and the penalty is multiplied by a scale factor, $C_{A}$, to control how stringently the penalty is enforced.

The second penalty prevents tethers from being positioned too close to, or outside, the mould boundary. 

We assign a minimum distance, $a_{\mathrm{min}}$ between tethers and mould edges (set here to 0.06$a$). This is also a soft limit. The penalty itself is calculated by measuring the shortest distance between the edge of tether $i$ and the boundary of the mould, $d_{i}$. 
\begin{equation}
p_{t} = C_{t}\sum_{i}{\rm max}\left(1-\frac{d_{i}}{a_{\mathrm{min}}},0\right).
\label{eqn:tetherpenalty}
\end{equation}
Where this penalty is multiplied by a scale factor denoted $C_{t}$. 

Then the total fitness to be maximised is,
\begin{equation}
F = \langle \svec_{x}\rangle - p_{A}-p_{t},
\end{equation}
where both penalties are subtracted from the initial fitness value to reduce the overall ranking of moulds with penalties. For this optimisation, we set $C_{A}=C_{t}=10$.

\subsubsection{Uniform Density}

The second optimisation goal is to generate tissue with an area of uniform cell density within a specific region of the mould area. We set this area as a square at the centre of the  canvas with dimensions $a/2\times a/2$. For the fitness value calculation we primarily make use of the cell density $\Rho$ field obtained from RAPTOR. Our measure of uniform density is calculated by taking the mean of the difference between values in this central region and the mean density, $\overline{P(A_{c})}$, which we shall refer to as the uniformity:
\begin{equation}
f_{\Rho} = C_{\Rho}\frac{\sum^{N}_{i=0} |P(i)-\overline{P(A_{c})}|}{N} 
\label{eqn:unform_density_fit}
\end{equation}
where N is the total number of pixels in the central area and $P \left ( i \right )$ is the density value of each pixel. $f_{\Rho}$ is minimised to find candidates with less deviation from the mean density and therefore a more uniform tissue. We set $C_{\Rho}=10$.

Three penalties are applied to this optimisation. Penalties (1) and (2) ensure minimum mould area and tether separation, as defined in Equations \ref{eqn:areapenalty} and \ref{eqn:tetherpenalty}. 

The third penalty constrains maximum mould area to reduce uncontrolled regions of density outside the target area.
\begin{equation}
p_{M} = C_{M}{\rm max}\left(\frac{A_{\mathrm{mould}}}{A_{\mathrm{central}}} - 1.1,0\right)
\end{equation}
where $A_{\mathrm{mould}}$ is the total area of the generated mould and $A_{\mathrm{central}}$ is the central evaluated area.

Here, the total fitness,
\begin{equation}
F = f_{\Rho} + p_{A} + p_{t} + p_{M},
\end{equation}
is minimised so penalties are added the the fitness value. We set scale factors to $C_{M}=C_{A}=C_{t}=C_{\rho}=10$.

\section{Results}
\label{sec:results}

We have tested three cases for mould design optimisation. These are maximum alignment with a low cell-matrix interaction ($\Delta=0.25$), maximum alignment with high cell-matrix interaction ($\Delta= 0.75$) and uniform density with low cell-matrix interaction ($\Delta =0.25$). In each case, after the optimisation routine was complete, we took a selection of the best mould designs (i.e. the highest fitness moulds in the final population) and used them as inputs to CONDOR simulations to compare tissue organisations and fitnesses from the full biophysical simulation results against the RAPTOR predictions and the fitnesses generated by the genetic algorithm.

\subsection{Maximum $x$-axis alignment for low cell-matrix interaction}

\begin{figure}
\centering
{
\includegraphics[width=0.47\textwidth]{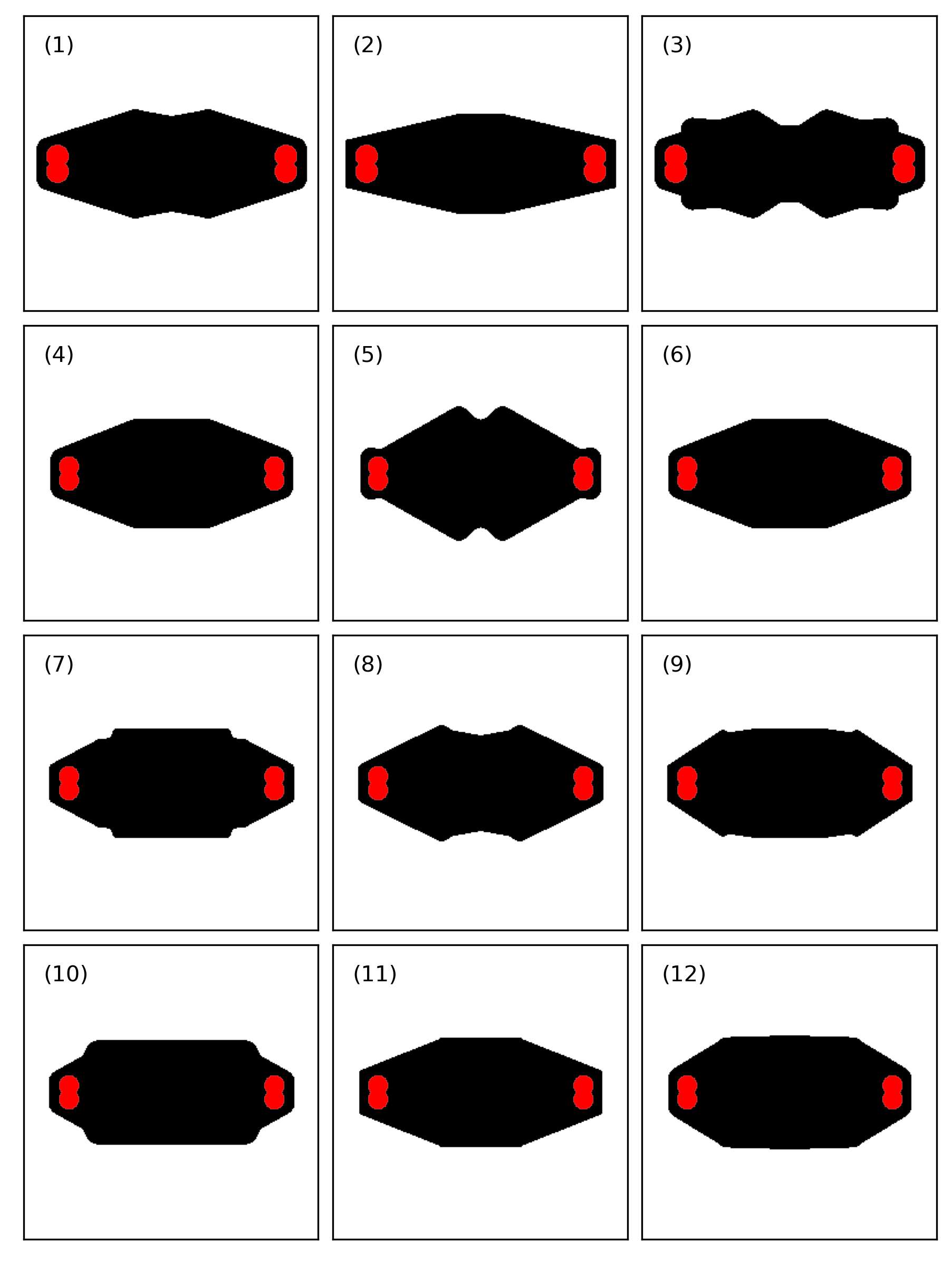}
\caption{The twelve best moulds for promoting maximum alignment in tissue, designed by the genetic algorithm for tissue with a low cell-matrix interaction, $\Delta$ (such as glial tissue).}
\label{fig:max_alignment_low_delta_mould_examples}
}
\end{figure}

Figure~\ref{fig:max_alignment_low_delta_mould_examples} shows the top $12$ mould designs produced by the genetic algorithm for maximum alignment at low cell-matrix interaction. As the result of crossover operations in the algorithm, specific successful design features have propagated through the population. Tether sizes and positions are very similar in each candidate, in each case consisting of a pair of large joint tethers at either end of each mould's longer axis; this indicates the particular tether arrangement produces the best results out of all possibilities and has subsequently propagated through the majority of the population. Moulds have more varied designs, around the common theme of an approximately diamond-shaped configuration with either flattened corners or concave notches in the centre. We note that the population contains some cases with duplicate mould designs and equal rankings, and only a single example of such cases is shown. 

\begin{figure}
\centering
{
\includegraphics[width=0.47\textwidth]{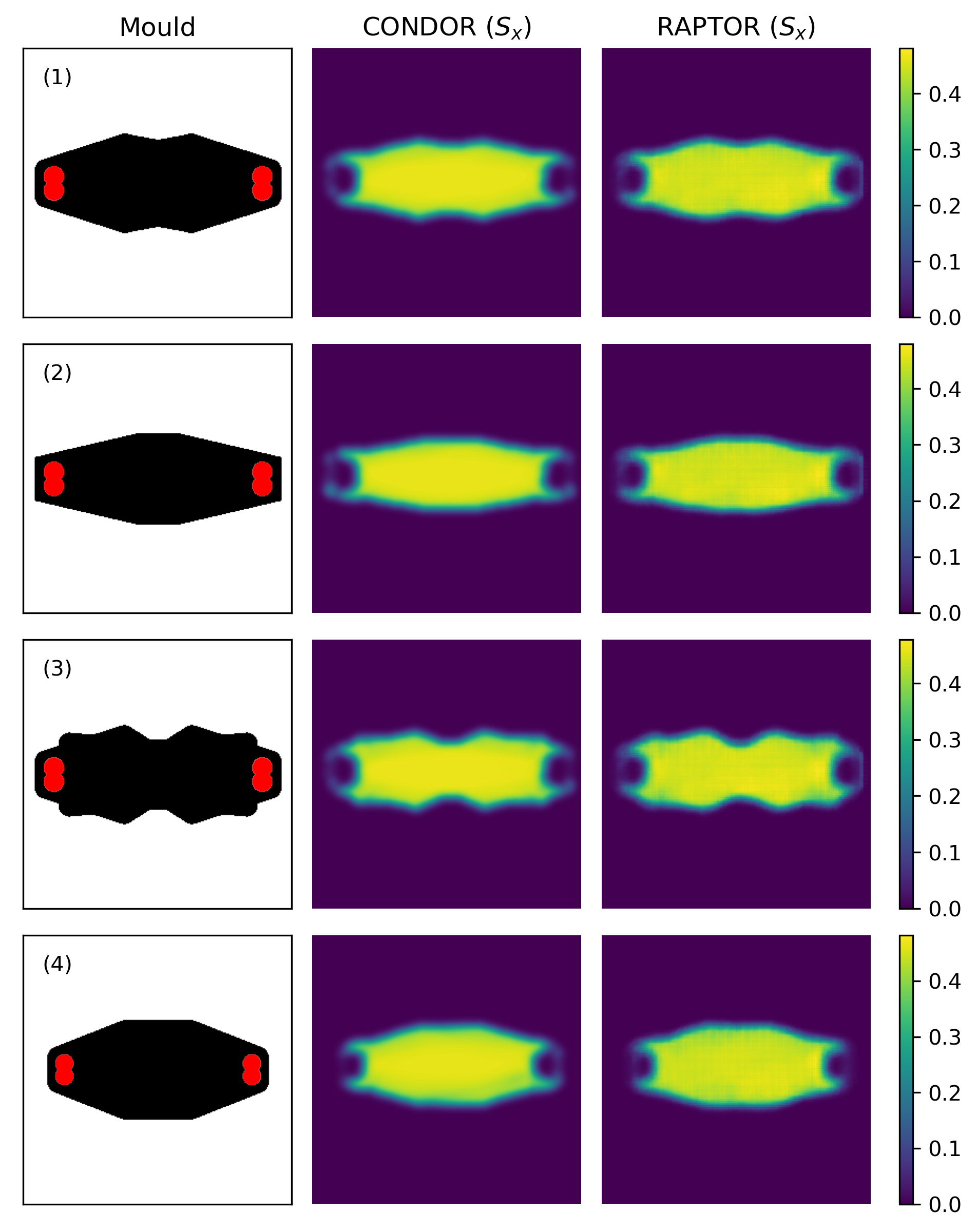}
\caption{Cell alignment, $S_{x}$, along the $x$ direction (across the page) for four moulds predicted by the genetic algorithm to have the highest maximum alignment when $\Delta$ is small (with the same ranking as Fig. \ref{fig:max_alignment_low_delta_mould_examples}).  Results can be seen for both CONDOR (left) and RAPTOR (right). Large regions of highly aligned tissue are predicted.}
\label{fig:max_alignment_low_delta_results}
}
\end{figure}

In Figure~\ref{fig:max_alignment_low_delta_results}, we show the results for the x-axis cell alignment field, $S_{x}$, from both the RAPTOR predictions as well as CONDOR simulations for the four highest ranked mould designs from Figure~\ref{fig:max_alignment_low_delta_mould_examples}. The corresponding fitness values for each mould are indicative of high total alignment of cells in both the simulations and predictions, are shown in Table~\ref{tab:max_alignment_low_delta_fit_vals}. These show that the predicted alignments from RAPTOR are high for each of the mould examples, with high alignment also seen in the CONDOR biophysical simulations (which predict marginally lower alignments); in every case, the majority of cells particularly in the central bulk of the cell matrix are approximately uniformly aligned.

\begin{table}
    \centering
    \begin{tabular}{c|c|c}
        rank & \multicolumn{2}{c}{alignment (fitness)} \\
        & RAPTOR & CONDOR\\
        \hline
        $1$ & $0.89989$ & $0.88327$\\
        $2$ & $0.89364$ & $0.88503$\\
        $3$ & $0.89214$ & $0.87153$\\
        $4$ & $0.88972$ & $0.87514$\\
    \end{tabular}
    \caption{Comparisons between the genetic algorithm fitness scores and equivalent calculated fitness values from CONDOR simulation results for the candidates with low $\Delta$ shown in Figure~\ref{fig:max_alignment_low_delta_results}. Alignment and fitness scores are the same because the penalties are zero once the moulds are optimised.}
    \label{tab:max_alignment_low_delta_fit_vals}
\end{table}

\subsection{Maximum $x$-axis alignment for high cell-matrix interaction}

\begin{figure}
\centering
{
\includegraphics[width=0.47\textwidth]{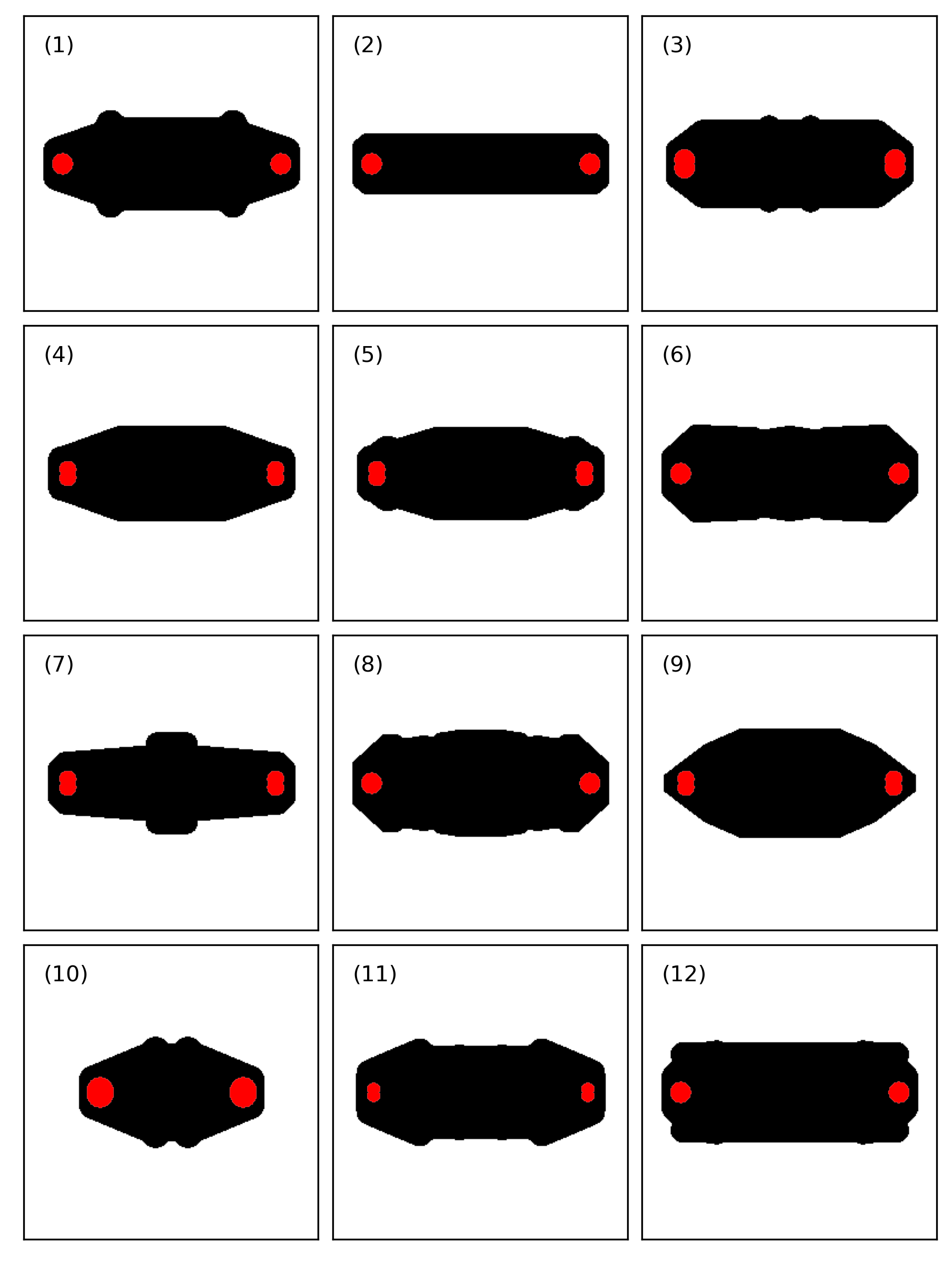}
\caption{The twelve best moulds for promoting maximum alignment in tissue, designed by the genetic algorithm for tissue with a high cell-matrix interaction, $\Delta$ (such as tendon or muscle).}
\label{fig:max_alignment_high_delta_mould_results}
}
\end{figure}

Figure~\ref{fig:max_alignment_high_delta_mould_results} shows the top $12$ mould design candidates produced by the genetic algorithm for maximum alignment at high $\Delta$. Again, successful mould design features including tether configurations have propagated through the population. Single tether configurations, or occasionally adjoined pairs. can be found at the ends of the mould. The mould designs are more elongated, with a shape closer to an approximately rectangular shape rather than the diamond shapes found in the low $\Delta$ cases.

\begin{figure}
\centering
{
\includegraphics[width=0.47\textwidth]{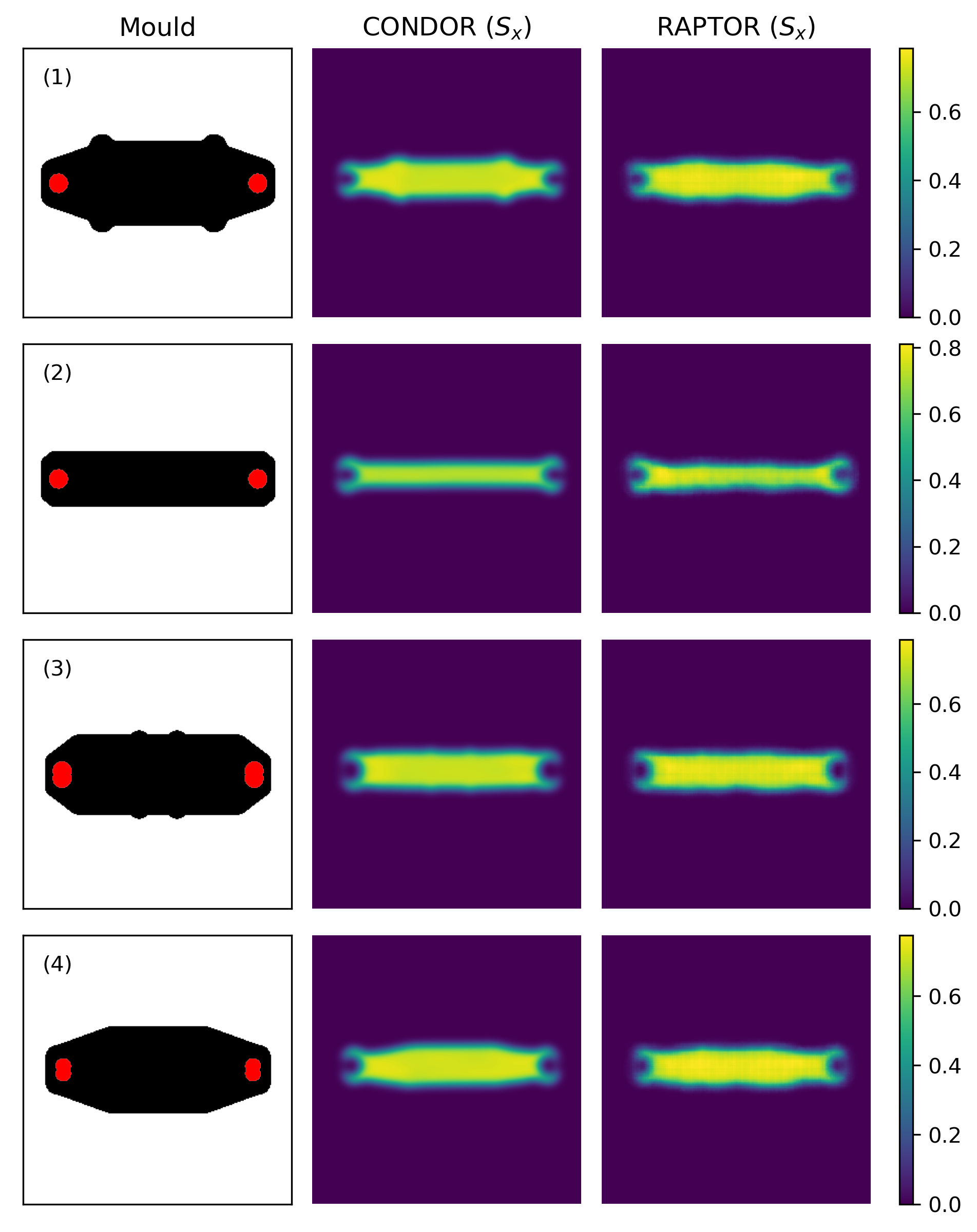}
\caption{Cell alignment, $S_{x}$, along the $x$ direction (across the page) for four moulds designs from the genetic algorithm to have the highest maximum alignment when $\Delta$ is large (with the same ranking as Fig. \ref{fig:max_alignment_high_delta_mould_results}).  Results can be seen for both CONDOR (left) and RAPTOR (right). Narrow regions of highly aligned tissue are predicted.}
\label{fig:max_alignment_high_delta_mould_examples}
}
\end{figure}

Again, we show comparisons between the RAPTOR predictions and CONDOR results for the four top (non-repeated) mould examples in Figure~\ref{fig:max_alignment_high_delta_mould_examples}. Where there is a high cell-matrix interaction ($\Delta=0.75$), the resultant cell distributions are much more contracted relative to the mould shapes, creating highly elongated final configurations. 

There is very good visual agreement in terms of the cell distributions between CONDOR and RAPTOR. The CONDOR simulations predict slightly lower total alignment than the RAPTOR predictions, Table~\ref{tab:max_alignment_high_delta_fit_vals}. The average alignments are lower in the high $\Delta$ case than the low $\Delta$ case, but still above $0.8$.

\begin{table}
    \centering
    \begin{tabular}{c|c|c}
        Candidate & RAPTOR alignment & CONDOR alignment\\
        $1$ & $0.84797$ & $0.81634$\\
        $2$ & $0.84575$ & $0.81938$\\
        $3$ & $0.84364$ & $0.81794$\\
        $4$ & $0.83961$ & $0.81340$\\
    \end{tabular}
    \caption{Comparisons between the genetic algorithm fitness scores and equivalent calculated fitness values from CONDOR simulation results for the candidates shown in Figure~\ref{fig:max_alignment_high_delta_mould_examples}.}
    \label{tab:max_alignment_high_delta_fit_vals}
\end{table}

\subsection{Uniform Density}

\begin{figure}
\centering
{
\includegraphics[width=0.47\textwidth]{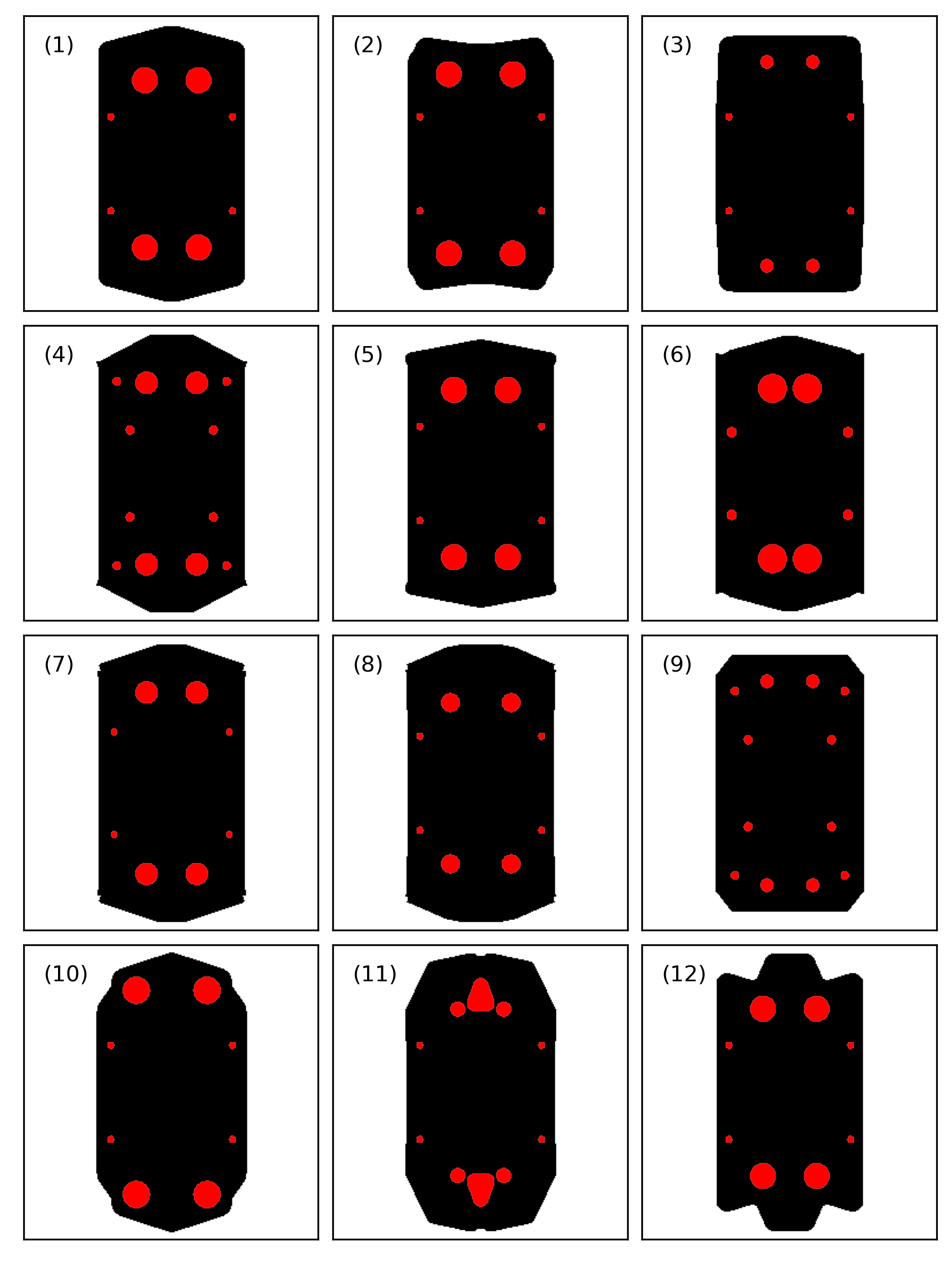}
\caption{The twelve best mould designs produced by the genetic algorithm optimisation for uniform density.}
\label{fig:uniform_density_mould_examples}
}
\end{figure}

The top $12$ mould designs produced by the genetic algorithm for uniform density are shown in Figure~\ref{fig:uniform_density_mould_examples}. The fitness function and penalty calculations are very different in this case, so the genetic algorithm has produced mould designs with very different configurations than those suitable for maximum alignment. The mould shapes in this case are almost rectangular, mostly with large tethers situated close to the corners, with additional smaller tethers positioned relatively close to the mould boundaries, surrounding the central region of the mould where the uniform density measure is evaluated.

\begin{figure}
\centering
{
\includegraphics[width=0.47\textwidth]{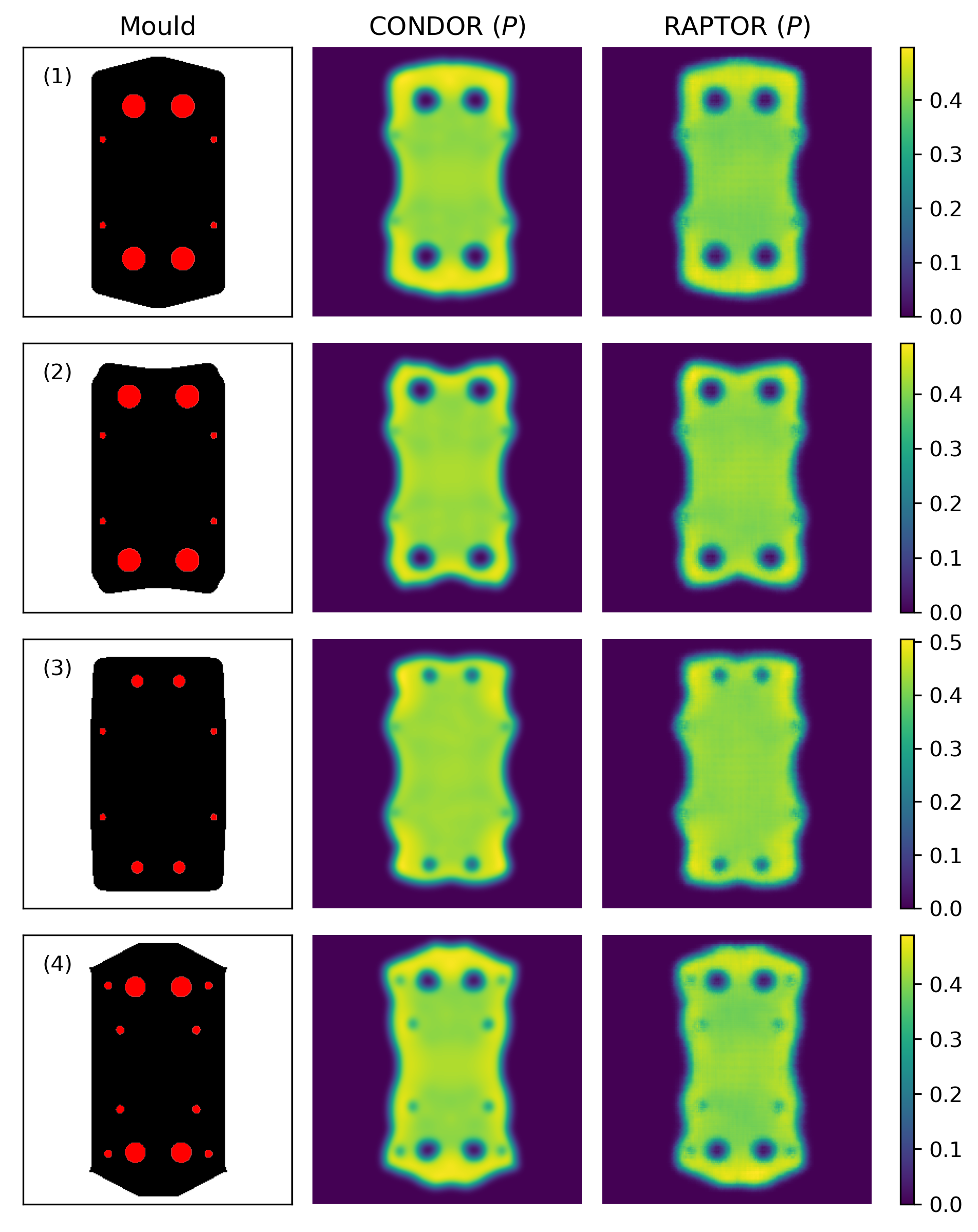}
\caption{CONDOR and RAPTOR predictions for tissue growth in the four best moulds designs that have been optimised for uniform density (numbered corresponding to \ref{fig:uniform_density_mould_examples}).}
\label{fig:uniform_density_low_delta_results}
}
\end{figure}

CONDOR results and RAPTOR predictions for cell density for the four best (non-repeating) mould designs are shown in Figure~\ref{fig:uniform_density_low_delta_results}. As with the previous optimisation cases, there is good visible agreement between CONDOR and RAPTOR results, though the RAPTOR predictions have a minor decrease in density horizontally at the centre. We also show the final fitness values (including penalties), along with fitness values without penalties applied (i.e. using Equation~\ref{eqn:unform_density_fit}) in Table~\ref{tab:uniform_density_low_delta_fit_vals}.

\begin{table}
    \centering
    \begin{tabular}{c|c|c|c}
        rank & fitness & \multicolumn{2}{c}{uniformity} \\
        & (with penalities) & RAPTOR & CONDOR \\
        \hline
        $1$ & $3.99635$ & $1.17267$ & $1.28889$\\
        $2$ & $4.01016$ & $1.26990$ & $1.35853$\\
        $3$ & $4.02117$ & $1.20547$ & $1.29043$\\
        $4$ & $4.02452$ & $1.16811$ & $1.27174$\\
    \end{tabular}
    \caption{Comparisons between the genetic algorithm fitness scores (including penalties), and uniform density values from Eqn.~\ref{eqn:unform_density_fit} from RAPTOR predictions and from CONDOR simulation results for the candidates shown in Fig.~\ref{fig:uniform_density_low_delta_results}.}
    \label{tab:uniform_density_low_delta_fit_vals}
\end{table}

\section{Discussion and conclusions}
\label{sec:conclusions}

In summary, we have developed a framework using evolutionary strategies to create mould and tether designs for cultured tissues with desirable properties such as cell alignment or uniform density. We combined an evolutionary optimisation strategy with rapid predictions made possible by RAPTOR \citep{andrews2023}, our machine-learning tissue organisation model. The resulting framework is an artificial intelligence approach to designing tethered moulds for engineered tissue.

One of the big challenges of developing cultured tissue growth strategies is that the tissue is a complex system that self-organises in a bottom-up process, whereas human-led design is primarily top-down. The design framework developed here allows top-down design principles to be applied for the tissue self-organisation, for a design process that can be be completed with AI relatively quickly. The primary human task is defining a suitable fitness function to identify good solutions.

We have tested three specific design scenarios: maximised cell alignment for low cell-matrix interaction, $\Delta$ (a less contractile tissue), maximum alignment for high $\Delta$ (more contractile tissue), and uniform cell density within a set area. Results for mould designs produced in each case were compared to the full CONDOR simulations to check for the risk that outliers in the machine learning predictions were found by the genetic algorithm. In each of the three cases considered, the genetic algorithm created mould designs that were optimal with respect to both RAPTOR and CONDOR.

A future extension could involve the direct use of CONDOR simulations as input to the fitness functions within the genetic algorithm. The benefit of using the biophysical model directly is that different types of growth methods are easier to implement in CONDOR (for example, to investigate 3D scaffolds, a separate machine learning solution would need to be trained). The downside is that direct use of CONDOR with the genetic algorithm would require a huge computational resource as the $\sim 100,000$ simulations (for 100 generations with 1000 genomes, each taking around 2 core days) leads to calculations that would take over 100 core years. Therefore, the speed up from machine learning is currently needed. We anticipate that compute time will eventually decrease to tractable levels due to growth in the number of logical cores in CPUs, since genetic algorithms operate in parallel.

The complex process of design using our hybrid machine learning and genetic algorithm approach takes under an hour when using a commercial laptop, and generates designs similar to those of detailed biophysical simulations. Around 100 core weeks of CONDOR calculation is needed to generate the training data but no further computation is needed following those calculations (noting that the training data can be used to train improved machine learning approaches as networks and strategies become more sophisticated). A limitation of machine learning is that individual training sets are currently needed for different problems, e.g. tethered moulds, spheroids, scaffolds (although as machine learning becomes more generalised it may be possible to develop a general machine learning approach). This compares effectively with the cost of lab-based design, where it can take months to develop designs for a single new mould. Work is ongoing to test our mould designs using laboratory cultivated tissues.

Another future extension involves investigation of alternative 3D scaffolding strategies and bioprinting strategies. These would require the training of new machine learning solutions. The development of new training data sets is expected to be faster than implementing CONDOR directly into the genetic algorithm. Such 3D extensions would open the possibility of AI design of scaffolds and bioprinting strategies for regenerative medicine and cultivated meat applications.

\bibliography{references}

\end{document}